\documentclass[paper]{JHEP3}
\pdfoutput=1
\usepackage{amssymb}
\usepackage{amsmath}
\usepackage{cite}
\usepackage{epsfig}
\def\nc{\newcommand}
\nc{\half}{\frac{1}{2}}
\nc{\shalf}{\ensuremath{\textstyle \frac{1}{2}}}

\nc{\deldag}{\mathbin{\partial\mkern-10.5mu\big/}}
\nc{\deldagss}{\mathbin{\partial\mkern-10.5mu/}}
\nc{\kdag}{\mathbin{k\mkern-10mu\big/}}
\nc{\skdag}{\mathbin{k\mkern-10mu/}}
\nc{\udag}{\mathbin{u\mkern-10mu\big/}}
\nc{\kdagss}{\mathbin{k\mkern-10mu/}}
\nc{\Pdag}{\mathbin{P\mkern-10mu\big/}}
\nc{\pp}{{\scriptscriptstyle ||}}
\nc{\stwo}{{\scriptscriptstyle 2}}
\nc{\pham}{{\phantom{-}}}

\def\lsim{\mathrel{\raise.3ex\hbox{$<$\kern-.75em\lower1ex\hbox{$\sim$}}}}
\def\gsim{\mathrel{\raise.3ex\hbox{$>$\kern-.75em\lower1ex\hbox{$\sim$}}}}

\def\Slashnew#1{#1\kern-0.55em\raise.05ex\hbox{/}}
\def\slashnew#1{#1\kern-0.5em\raise.05ex\hbox{{$\scriptstyle /$}}}

\def\ie{{\em i.e. }}

\nc{\beq} {\begin{equation}}
\nc{\eeq} {\end{equation}}
\nc{\beqa}{\begin{eqnarray}}
\nc{\eeqa}{\end{eqnarray}}

\def\Slashnew#1{#1\kern-0.55em\raise.05ex\hbox{/}}
\def\slashnew#1{#1\kern-0.5em\raise.05ex\hbox{{$\scriptstyle /$}}}
\def\emph#1{{\em #1}}
\def\hepph#1{hep-ph/#1}

\title{Flavour-coherent propagators and Feynman rules: Covariant cQPA formulation}

\author{Matti Herranen$^{a,}$\footnote{Alexander-von-Humboldt Fellow}\,\,,
        Kimmo Kainulainen$^{b,c}$ and
        Pyry Matti Rahkila$^{b,c}$ 
        \\
	\\ $^a$Institut f\"ur Theoretische Teilchenphysik und Kosmologie,
	\\ RWTH Aachen University, D--52056 Aachen, Germany
        \\ $^{b}$ Department of Physics, P.O.~Box 35 (YFL), 
        \\ FIN-40014 University of Jyv\"askyl\"a, Finland, 
        \\ $^{c}$ Helsinki Institute of Physics, P.O.~Box 64, 
  	\\ FIN-00014 University of Helsinki, Finland.\\
        \\e-mail: \email{herranen@physik.rwth-aachen.de,   
                         kimmo.kainulainen@jyu.fi, 
                         pyry.rahkila@jyu.fi}}

\abstract{We present a simplified and generalized derivation of the flavour-coherent propagators 
and Feynman rules for the fermionic kinetic theory based on coherent quasiparticle approximation 
(cQPA) \cite{HKR1,HKR2,HKR3,Glasgow,Thesis_Matti,HKR4,FHKR}.
The new formulation immediately reveals the composite nature of the cQPA Wightman function as a 
product of two spectral functions and an effective two-point interaction vertex, which contains all 
quantum statistical and coherence information. We extend our previous work to the case 
of nonzero dispersive self-energy, which leads to a broader range of applications.
By this scheme, we derive flavoured kinetic equations for local 2-point functions 
$S^{<,>}_\mathbf{k}(t,t)$, which are reminiscent of the equations of motion for the density 
matrix. We emphasize that in our approach all the interaction terms are derived from first 
principles of nonequilibrium quantum field theory.} 

\preprint{TTK-11-36}

\begin{document}

%%%%%%%%%%%%%%%%%%%%%%%%%%%%%%%%%%%%%%%%%%%%%%%%%%%%%%%%%%%%%%%%%%%%%%%%%%%%%%%%
%%%%%%%%%%%%%%%%%%%%%%%%%%%%%%%%%%%%%%%%%%%%%%%%%%%%%%%%%%%%%%%%%%%%%%%%%%%%%%%%
%
\section{Introduction}
%
%%%%%%%%%%%%%%%%%%%%%%%%%%%%%%%%%%%%%%%%%%%%%%%%%%%%%%%%%%%%%%%%%%%%%%%%%%%%%%%%
%%%%%%%%%%%%%%%%%%%%%%%%%%%%%%%%%%%%%%%%%%%%%%%%%%%%%%%%%%%%%%%%%%%%%%%%%%%%%%%%

Coherence phenomena play an important role in many interesting problems of particle physics and 
cosmology. Well known examples include neutrino oscillations~\cite{neutrino_osc1,neutrino_osc2}, electroweak baryogenesis
 \cite{EWBG,ClassForce,ClassForceomat,SemiClassSK,PSW,EWBG_mixing,Cirigliano}, coherent baryogenesis 
\cite{coherent_bg,HKR4}, leptogenesis \cite{leptogenesis,resonant_lepto,flavour-CTP_lepto,ABDM10} 
and out-of-equilibrium particle production \cite{particle_prod}. Recently, 
a novel kinetic theory approach which includes both flavour- and particle-antiparticle coherence 
effects was introduced in refs.~\cite{HKR1,HKR2,HKR3,Glasgow,Thesis_Matti,HKR4,FHKR}. This formalism, 
based on the coherent quasiparticle approximation (cQPA), is in principle applicable to all the problems 
mentioned above. Other recent works related to quantum coherence and decoherence in quantum field theory 
include {\em e.g.} \cite{Cirigliano,coherence_QFT}.

cQPA scheme is based on the Schwinger-Keldysh formalism of nonequilibrium quantum field theory 
\cite{SK-formalism} with the approximations of weak interactions and slowly varying background field 
({\em e.g.} expansion of the Universe). It provides an extension to standard kinetic theory with quantum
 Boltzmann equations (see {\em e.g.} refs.~\cite{kinetic,PSW,CalHu08}) in that in cQPA the 2-point correlators
 are {\em not} assumed to be nearly time-translation invariant, which allows new oscillatory solutions
 mediating non-local
 quantum coherence \cite{HKR1,HKR2,HKR3}. Because of the rapid oscillations at quantum scales $\sim k$,
 the collision integrals involving these coherence solutions need to be resummed to all orders of the
 gradient expansion in Wigner representation \cite{Glasgow,Thesis_Matti,HKR4}. This resummation procedure
 ultimately leads to an extended set of Feynman rules for coherent systems, as well as coherent quantum
 Boltzmann equations for enlarged set of distribution functions, formulated in \cite{HKR4} for single
 fermionic and scalar fields. The generalization of the formalism to the systems with multiple mixing
 fermionic and scalars fields was presented in \cite{FHKR}.

In this work, we present a new, more elegant and general, flavour 
covariant formulation of the fundamental cQPA Wightman functions and the perturbative Feynman rules 
for the theory. The cQPA Wightman function is found to be a composite operator, schematically of the
 form $iS^{<,>}_{ab} = {\cal A}_a F^{<,>}_{ab}(t,t){\cal A}_b$, where the spectral functions ${\cal A}$
 project $S^{<,>}_{ab}$ to the mixing states labeled by $a$ and $b$, and the local 2-point vertex function
 $F^{<,>}_{ab}(t,t)$ contains all associated statistical information. The composite structure of the
 (statistical) propagator was also observed in refs.~\cite{GreLeu98,ABDM09} for nonmixing scalar and in
 \cite{ABDM10} for fermionic fields coupled to thermal bath. In these works, the local correlators
 $F^{<,>}$ were enscribing the initial conditions of the system, whereas in our approach $F^{<,>}_{ab}(t,t)$
 are the dynamical variables of interest. This (derived) form can be viewed as a generalization to coherent
 systems of the well known Kadanoff-Baym ansatz \cite{KadBay62}, which projects the propagator to a single
 mass shell: $iS^{<,>}_{aa} \sim 2{\cal A}_an_{a}$, where $n_a$ is particle number in the state $a$.

We generalize our previous work also by including the dispersive (hermitian) part of the self-energy in
 the dispersion relations, which extends our formalism to treatment of quasi-excitations in plasmas.
 This is important for example for a consistent approach to neutrino oscillations phenomena and thermal
 resonant leptogenesis
 \cite{resonant_lepto}, as well as for the electroweak baryogenesis. We note that the final flavoured
 kinetic equations for local 2-point functions $S^{<,>}_\mathbf{k}(t,t)$, resulting from our approach,
 are very reminiscent of the equations of motion for the (flavoured) density matrix in the standard 
 density matrix approach (see {\em e.g.} \cite{neutrino_osc1}).
 We want to emphasize, however, that in our approach all the interaction terms are derived from first
 principles of nonequilibrium QFT. Moreover, these equations are local in time-variable $t$, and they can
 be written as flavoured quantum Boltzmann equations for the (enlarged) set of on-shell distribution
 functions $f(\mathbf{k},t)$. 

The paper is organized as follows. In section \ref{sec:KB_cQPA}, we briefly review the Kadanoff-Baym equations
 and the cQPA scheme. In section \ref{sec:propagators}, we present the flavour covariant derivation of the
 cQPA propagators, and write down the kinetic equations in section \ref{sec:kinetic_eq1}.
 In section \ref{sec:feynman}, we show the covariant formulation of the cQPA Feynman rules, and in section
 \ref{sec:kinetic_eq2} we use these rules to write down the final kinetic equations for the local correlators.
 In this section, we also present a parametric estimate for the error resulting from our approximations.
 Finally, section \ref{sec:conclusions} contains our conclusions.

%%%%%%%%%%%%%%%%%%%%%%%%%%
%%%%%%%%%%%%%%%%%%%%%%%%%%
\section{KB-equations and cQPA}
\label{sec:KB_cQPA}
%%%%%%%%%%%%%%%%%%%%%%%%%%
%%%%%%%%%%%%%%%%%%%%%%%%%%

We consider a spatially homogeneous and isotropic flavour-mixing fermionic system with a complex and possibly 
time-dependent mass matrix $m$. The complete Kadanoff-Baym (KB) equations \cite{KadBay62}
for the 2-point correlation functions in two-time representation can be written as (cf. \cite{HKR2})
\begin{align}
\big(i\gamma^0 \partial_{t_1} - \mathbf{k}\cdot\mathbf{\gamma} - m_h(t_1) 
- i \gamma^5 m_a(t_1)\big)S^{r,a}_\mathbf{k}(t_1,t_2)& 
\nonumber\\[1mm]
- \int {\rm d} t^\prime\,\Sigma^{r,a}_\mathbf{k}(t_1,t^\prime) S^{r,a,}_\mathbf{k}(t^\prime,t_2)&
= \delta(t_1-t_2)\,,
\label{KB_eq1}
\\[4mm]
\big(i\gamma^0 \partial_{t_1} - \mathbf{k}\cdot\mathbf{\gamma} - m_h(t_1) 
- i \gamma^5 m_a(t_1)\big)S^{<,>}_\mathbf{k}(t_1,t_2)& 
\nonumber\\[1mm]
- \int {\rm d} t^\prime\,\big(\Sigma^h_\mathbf{k}(t_1,t^\prime)S^{<,>}_\mathbf{k}(t^\prime,t_2)&
\nonumber\\
+ \,\Sigma^{<,>}_\mathbf{k}(t_1,t^\prime)S^{h}_\mathbf{k}(t^\prime,t_2)\big)& 
=  \pm {\cal C}^{\rm coll}_\mathbf{k}(t_1,t_2)\,.
\label{KB_eq2}
\end{align}
where $m_h=(m + m^\dagger)/2$ and $m_a = (m - m^\dagger)/(2i)$, $S^{r,a}$ are the standard 
(retarted and advanced) pole propagators, $iS^<(u,v) = \langle \bar\psi(v)\psi(u) \rangle$ 
and $iS^>(u,v) = \langle \psi(u) \bar\psi(v)\rangle$
are the Wightman functions, and $S^h = (S^r+ S^a)/2$, with similar definitions for the 
corresponding self-energies $\Sigma$.\footnote{Note that our definitions of $S^<$ and $\Sigma^<$ 
differs by sign from the more standard convention.} The $\pm$ sign in Eq.~(\ref{KB_eq2}) refers to 
$<$ or $>$ component equations, respectively. The collision term is given by
\begin{equation}
{\cal C}^{\rm coll}_\mathbf{k}(t_1,t_2) = \frac{1}{2}\int {\rm d} t^\prime 
                             \big(\Sigma^>_\mathbf{k}(t_1,t^\prime) S^<_\mathbf{k}(t^\prime,t_2)
                                  -\Sigma^<_\mathbf{k}(t_1,t^\prime) S^>_\mathbf{k}(t^\prime,t_2)\big)\,,
\label{collintegral}
\end{equation}
and the correlators and self-energies in {\em two-time representation} are defined as
\begin{equation}
F_\mathbf{k}(t_1,t_2) \equiv \int {\rm d}^{\,3}(\mathbf{x_1} - \mathbf{x_2}) \, 
e^{-i\mathbf{k}\cdot(\mathbf{x_1} - \mathbf{x_2})} F(x_1,x_2)\,.
\label{prop_two-time}
\end{equation}
We will later on refer to {\em Wigner representation} as well, where the 2-point functions are defined
as
\begin{equation}
F(k,t) \equiv \int {\rm d} (t_1 - t_2) \, 
e^{i k_0 (t_1 - t_2)} F_\mathbf{k}(t_1,t_2)\,,
\label{prop_}
\end{equation}
where $k=(k_0, \mathbf{k})$ is the 4-momentum variable and $t \equiv (t_1 + t_2)/2$ is the average
 time-coordinate.

In cQPA we first expand the KB-equations (\ref{KB_eq1}-\ref{KB_eq2}) to the zeroth order
in time-gradients $\partial_t m(t)$ and in the scattering width $\Gamma = i(\Sigma^>+\Sigma^<)/2$
to obtain flavour-coherent propagators with (enlarged) singular quasiparticle phase-space 
structure. We also assume that the hermitian part of the self-energy, $\Sigma^h_\mathbf{k}$, 
does not depend on the dynamical correlators, and thus $\partial_t \Sigma^h_\mathbf{k}(t_1,t_2) 
\sim \partial_t m(t)$ can be neglected as well. The average-time dependence of the (dynamical)
 cQPA-propagators $S^{<,>}$ is found to be factorized to local correlators 
$S^{<,>}_\mathbf{k}(t,t)$ or equivalently to the set of on-shell distribution functions 
$f(\mathbf{k},t)$. To obtain the (first order) equations of motion for $S^{<,>}_\mathbf{k}(t,t)$, we 
then insert the dynamical cQPA-propagators back to Eq.~(\ref{KB_eq2}) and expand to (combined) 
first order\footnote{The dimensionless expansion parameters are roughly 
$\partial_t m/\omega_\mathbf{k}^2 $, $\Gamma_\mathbf{k} / \omega_\mathbf{k}$ and  
$\Sigma^h_\mathbf{k}/\omega_\mathbf{k}$. See section \ref{sec:error} below for further discussion.}
 in $\partial_t m$, $\Gamma$ and $\Sigma^h$. These equations are local in time-variable $t$,
 and they can be written as flavoured quantum Boltzmann equations for the distribution
 functions $f(\mathbf{k},t)$.

%%%%%%%%%%%%%%%%%%%%%%%%%%
%%%%%%%%%%%%%%%%%%%%%%%%%%
\section{Structure of propagators}
\label{sec:propagators}
%%%%%%%%%%%%%%%%%%%%%%%%%%
%%%%%%%%%%%%%%%%%%%%%%%%%%

%%%%%%%%%%%%%%%%%%%%%%%%%%%%%%%%%%%%%%%%%%%%%%%%%%%%%%%%%%%%%%%%%%%%%%%%%%%%%%%%
%
\subsection{Pole propagators and the spectral function}
%
%%%%%%%%%%%%%%%%%%%%%%%%%%%%%%%%%%%%%%%%%%%%%%%%%%%%%%%%%%%%%%%%%%%%%%%%%%%%%%%%

To the zeroth order in $\partial_t m$, the non-dynamical pole propagators $S^{r,a}$ are time-translation
 invariant and the KB-equations (\ref{KB_eq1}) in Wigner representation read\footnote{If the average-time
 derivatives $\partial_t S^{r,a}(k,t)$ would have been kept in Eq.~(\ref{KB-pole}), they would have
 eventually been found to vanish in the limit of vanishing $\partial_t m$ and $\partial_t \Sigma(k)$.}
\begin{equation}
\big(\kdag - m_h - i \gamma^5 m_a - \Sigma^h(k) \pm i \Gamma(k)\big) S^{r,a}(k) = 1\,,
\label{KB-pole}
\end{equation}
with a simple solution
\begin{equation}
S^{r,a}(k) = \frac{1}{\kdag - m_h - i \gamma^5 m_a - \Sigma^h(k) \pm i \Gamma(k)}\,,
\label{pole}
\end{equation}
where we have used $\Sigma^{r,a} = \Sigma^h \mp i \Gamma$ and we keep infinitesimal $\Gamma$. Even though
 the time-gradients $\partial_t m$ and $\partial_t \Sigma^h(k)$ are neglected here, $m$ and $\Sigma^h(k)$
 are understood to depend adiabatically on the (average-)time coordinate. Note 
that the masses $m_{h,a}$ and the self-energies $\Sigma^h$ and $\Gamma$ carry flavour indices, 
and thus the computation of the inverse matrices in Eq.~(\ref{pole}) can in general be tedious.
The spectral function is then formally given by  (we now take the limit $\Gamma \to 0$)
\begin{align}
{\cal A}(k) = \frac{i}{2}\big(S^{r}(k) - S^{a}(k)\big)
\equiv& \,\pi\,{\rm sgn}(k_0) \delta\big(\kdag - m_h - i \gamma^5 m_a - \Sigma^h(k)\big)
\nonumber\\
=& \,2 \pi \sum_i A_{i \mathbf{k}}\,\delta(k_0 - \tilde\omega_{i\mathbf{k}})\,,
\label{spectral}
\end{align}
where $A_{i \mathbf{k}}$ are matrices in Dirac and flavour indices, and $\tilde\omega_{i\mathbf{k}}$
 are the energies of the quasiparticle states, labeled by index $i$. 
We note that in two-time representation the spectral function (\ref{spectral}) satisfies 
the zeroth order equation 
\begin{equation}
\big(i\gamma^0 \partial_y  - \mathbf{k}\cdot\mathbf{\gamma} - m_h 
- i \gamma^5 m_a\big){\cal A}_\mathbf{k}(y)
- \int {\rm d} y^\prime\,\Sigma^h_\mathbf{k}(y-y^\prime){\cal A}_\mathbf{k}(y^\prime) = 0\,.
\label{spec_eq_two-time}
\end{equation}
where $y \equiv t_1 -t_2$ is the relative time-coordinate.
Moreover, from the singularity structure of Eq.~(\ref{spectral}) it clearly follows that
 ${\cal A}_\mathbf{k}(y)$ is analytic in $y$, and we can therefore expand the convolution integral
in Eq.~(\ref{spec_eq_two-time}) as
\begin{equation}
\int {\rm d} y^\prime\,\Sigma^h_\mathbf{k}(y-y^\prime){\cal A}_\mathbf{k}(y^\prime) 
= \int {\rm d} y^\prime\,\Sigma^h_\mathbf{k}(y^\prime){\cal A}_\mathbf{k}(y-y^\prime) 
= \sum_{n=0}^\infty \Sigma^h_{\mathbf{k}n}\,i^n \partial_y^n {\cal A}_\mathbf{k}(y)\,,
\end{equation}
where
\begin{equation}
\Sigma^{h}_{\mathbf{k}n} \equiv \frac{i^n}{n!} \int {\rm d} y\,y^n\, 
\Sigma^h_\mathbf{k}(y)\,.
\label{sigma_moments}
\end{equation}
Using this expansion in Eq.~(\ref{spec_eq_two-time}), and solving iteratively for
 $\partial_y {\cal A}_\mathbf{k}(y)$ neglecting the contributions of order $(\Sigma^h)^2$ and 
 higher,\footnote{This is consistent with the general procedure of QPA, where the $k_0$-poles of the spectral
 function (\ref{spectral}) are solved iteratively by using the lowest order solution for $k_0$
 inside $\Sigma^h(k_0,\mathbf{k})$.} we find
\begin{equation}
\partial_y {\cal A}_\mathbf{k}(y) = -i H_{\mathbf{k},{\rm eff}}{\cal A}_\mathbf{k}(y)
\qquad \Rightarrow \qquad
{\cal A}_\mathbf{k}(y) = e^{-i H_{\mathbf{k},{\rm eff}}y} {\cal A}_{\mathbf{k}0}\,,
\label{spec_evolution}
\end{equation}
where we denote ${\cal A}_{\mathbf{k}0} \equiv {\cal A}_{\mathbf{k}}(0)$ and the 
effective quasiparticle Hamiltonian is given by
\begin{equation}
H_{\mathbf{k},{\rm eff}} \equiv H_{\mathbf{k}0} + \int {\rm d} y \,\gamma^0 \Sigma^h_{\mathbf{k}}(y)\, 
e^{i H_{\mathbf{k}0}y}\,,
\end{equation}
where $H_{\mathbf{k}0} \equiv \mathbf{k}\cdot\mathbf{\alpha} + \gamma^0 m_h + 
i \gamma^0\gamma^5 m_a$ is the free Hamiltonian of the system. Note that we have not used the spectral sum-rule: 
${\cal A}_{\mathbf{k}}(0) = \int \frac{{\rm d}k_0}{2\pi} {\cal A}(k) = \frac{1}{2}\gamma^0 \delta_{ij}$, here,
 because in general it does not hold exactly in QPA with pole contributions only~\cite{sum-rule_QPA}. It is now trivial to show that
 the quasiparticle spectral function satisfying Eq.~(\ref{spec_eq_two-time}) has an important
 time-additivity property:
\begin{equation}
{\cal A}_\mathbf{k}(y_1 + y_2){\cal A}_{\mathbf{k}0}^{-1} = 
{\cal A}_\mathbf{k}(y_1){\cal A}_{\mathbf{k}0}^{-1}{\cal A}_\mathbf{k}(y_2){\cal A}_{\mathbf{k}0}^{-1}\,,
\label{spectral_additivity}
\end{equation}
such that 
\begin{equation}
U_\mathbf{k}(y) \equiv {\cal A}_\mathbf{k}(y){\cal A}_{\mathbf{k}0}^{-1}
 = e^{-i H_{\mathbf{k},{\rm eff}}y}
\label{time-evo_operator}
\end{equation}
 can be interpreted as a {\em time-evolution operator}. Note that $U_\mathbf{k}(y)$ is not necessarily 
 unitary in quasiparticle approximation due to possible breaking of the spectral sum-rule. However,
 if the sum-rule holds, it follows from the general hermiticity property: ${\cal A}^\dagger_\mathbf{k}(y) =
 \gamma^0 {\cal A}_\mathbf{k}(-y)\gamma^0$, that $H_{\mathbf{k},{\rm eff}}$ must be hermitian and the
 ``free'' (quasiparticle) evolution is unitary. We want to emphasize that the additivity property
 (\ref{spectral_additivity}) holds irrespective of the unitarity of $U_\mathbf{k}(y)$.
 This property is crucial for showing below that the dependence on the
 relative coordinate $y = t_1 -t_2$ and and the average coordinate $t=(t_1+t_2)/2$ of the Wightman functions
 $S^{<,>}$ factorizes neatly in quasiparticle approximation, leading to (enlarged) singular phase-space
 structure and local time-evolution equations for the corresponding on-shell distribution functions.

%%%%%%%%%%%%%%%%%%%%%%%%%%%%%%%%%%%%%%%%%%%%%%%%%%%%%%%%%%%%%%%%%%%%%%%%%%%%%%%%
%
\subsection{Wightman functions $S^{<,>}$}
%
%%%%%%%%%%%%%%%%%%%%%%%%%%%%%%%%%%%%%%%%%%%%%%%%%%%%%%%%%%%%%%%%%%%%%%%%%%%%%%%%

To the zeroth order in $\partial_t m$ and $\Gamma \sim \Sigma^{<,>}$, the KB-equation (\ref{KB_eq2}) for the Wightman functions:
\begin{equation}
\big(i\gamma^0 \partial_{t_1} - \mathbf{k}\cdot\mathbf{\gamma} - m_h 
- i \gamma^5 m_a\big)S^{<,>}_\mathbf{k}(t_1,t_2) 
%\nonumber\\
- \int {\rm d} y^\prime\,\Sigma^h_\mathbf{k}(t_1-t^\prime)S^{<,>}_\mathbf{k}(t^\prime,t_2) = 0\,,
\label{wightman_eq_two-time}
\end{equation}
is identical to Eq.~(\ref{spec_eq_two-time}) for the spectral function with the replacement $y \to t_1$,
 and $t_2$ acting as a parameter. Therefore, it has a factorized general solution (matrix product): 
$S^{<,>}_\mathbf{k}(t_1,t_2) = {\cal A}_\mathbf{k}(t_1) C^{<,>}_\mathbf{k}(t_2)$, 
where $t_2$-dependence can be fixed by using the hermiticity properties: 
$i S^{<,>\dagger}_\mathbf{k}(t_1,t_2) = \gamma^0 i S^{<,>}_\mathbf{k}(t_2,t_1)\gamma^0$ and 
${\cal A}^\dagger_\mathbf{k}(y) = \gamma^0 {\cal A}_\mathbf{k}(-y)\gamma^0$, to get
\begin{equation}
S^{<,>}_\mathbf{k}(t_1,t_2) = {\cal A}_\mathbf{k}(t_1) {\cal A}_{\mathbf{k}0}^{-1} 
S^{<,>}_\mathbf{k}(0,0){\cal A}_{\mathbf{k}0}^{-1}{\cal A}_\mathbf{k}(-t_2)\,.
\label{combined_eq}
\end{equation}
Eq.~(\ref{combined_eq}) combines the zeroth-order time evolution in the relative coordinate 
$y = t_1-t_2$ and in the average coordinate $t=(t_1+t_2)/2$. This general structure of the 
statistical propagator was also observed in \cite{GreLeu98,ABDM09} for nonmixing scalar and in
 \cite{ABDM10} for fermionic field coupled to thermal bath. 

To factorize $y$- and $t$-dependencies in Eq.~(\ref{combined_eq}), we use the additivity 
property (\ref{spectral_additivity}) and Eq.~(\ref{combined_eq}) for local correlators
 $S^{<,>}_\mathbf{k}(t,t)$ to get
\begin{equation}
S^{<,>}_\mathbf{k}(t_1,t_2) = {\cal A}_\mathbf{k}(y/2) {\cal A}_{\mathbf{k}0}^{-1} 
S^{<,>}_\mathbf{k}(t,t){\cal A}_{\mathbf{k}0}^{-1}{\cal A}_\mathbf{k}(y/2)\,.
\label{phase_space_eq}
\end{equation}
In Wigner representation Eq.~(\ref{phase_space_eq}) becomes a convolution integral:
\begin{align}
S^{<,>}(k,t) =& \int \frac{{\rm d}k_0^\prime}{\pi} {\cal A}(k_0^\prime,\mathbf{k}) 
{\cal A}_{\mathbf{k}0}^{-1} S^{<,>}_\mathbf{k}(t,t)
{\cal A}_{\mathbf{k}0}^{-1}{\cal A}(2 k_0 - k_0^\prime,\mathbf{k})
\nonumber\\
=& \, 2\pi \sum_{ij} A_{i \mathbf{k}}{\cal A}_{\mathbf{k}0}^{-1}S^{<,>}_\mathbf{k}(t,t)
{\cal A}_{\mathbf{k}0}^{-1} A_{j \mathbf{k}}\, \delta\Big(k_0 - 
\frac{1}{2}(\tilde\omega_{i\mathbf{k}} + \tilde\omega_{j\mathbf{k}})
\Big)
\,,
\label{phase_space_eq2}
\end{align}
where we have used Eq.~(\ref{spectral}) for the spectral function.
This structure of cQPA Wightman functions is one of the main results of this paper. The local 
correlators $S^{<,>}_\mathbf{k}(t,t)$ contain the full statistical information of the system, 
including phase-space particle number and (flavour) coherence distribution functions. 
We note that if ${\cal A}_{\mathbf{k}0}^{-1}{\cal A}(k)$ commutes with 
${\cal A}_{\mathbf{k}0}^{-1}S^{<,>}_\mathbf{k}(t,t)$ for all $k_0$, the time-additivity 
property (\ref{spectral_additivity}) yields 
$S^{<,>}(k,t) = \\ {\cal A}(k) {\cal A}_{\mathbf{k}0}^{-1} S^{<,>}_\mathbf{k}(t,t)$. Moreover, 
in the absence of the dispersive self-energy $\Sigma^h$ and the coherence contributions (see
Eqs.~(\ref{free_spectral}-\ref{full_wightman_diag}) below), we recover the
standard form of {\em the Kadanoff-Baym ansatz}: $iS^{<,>}_{ii}(k,t) \propto 2 f^{<,>}_{ii}(k,t){\cal A}_{ii}(k)$,
 where $f^{<,>}_{ii}(k,t)$ are the phase-space distribution functions of flavour state $i$. Based on
 these considerations, Eq.~(\ref{phase_space_eq2}) can be viewed as a generalization of the
 Kadanoff-Baym ansatz to systems including flavour- and particle-antiparticle coherence.

We also note that the composite structure of the Wightman function (\ref{phase_space_eq2}) 
involves cross-couplings between different flavour-components of $S^{<,>}(k,t)$ and 
$S^{<,>}_{\mathbf{k}}(t,t)$, if the spectral function is not flavour diagonal.\footnote{Note that the
 cross-couplings always vanish in $k_0$-integration, reducing Eq.~(\ref{phase_space_eq2}) to trivial
 identity for local correlators.}
For example, in the case of two-flavour mixing we have:
\begin{align}
S^{<,>}_{12}(k,t) = 2 \int \frac{{\rm d}k_0^\prime}{2\pi}\bigg[ &\big({\cal A}(k_0^\prime,\mathbf{k}) {\cal A}_{\mathbf{k}0}^{-1}\big)_{11} S^{<,>}_{\mathbf{k},12}(t,t)\big({\cal A}_{\mathbf{k}0}^{-1}{\cal A}(2 k_0 - k_0^\prime,\mathbf{k})\big)_{22}
\nonumber\\
+ &\big({\cal A}(k_0^\prime,\mathbf{k}) {\cal A}_{\mathbf{k}0}^{-1}\big)_{11} S^{<,>}_{\mathbf{k},11}(t,t)\big({\cal A}_{\mathbf{k}0}^{-1}{\cal A}(2 k_0 - k_0^\prime,\mathbf{k})\big)_{12}
\nonumber\\
+ &\big({\cal A}(k_0^\prime,\mathbf{k}) {\cal A}_{\mathbf{k}0}^{-1}\big)_{12} S^{<,>}_{\mathbf{k},22}(t,t)\big({\cal A}_{\mathbf{k}0}^{-1}{\cal A}(2 k_0 - k_0^\prime,\mathbf{k})\big)_{22} \bigg]\,,
\label{phase_space_eq3}
\end{align}
and similarly for other components $S^{<,>}_{ij}(k,t)$.

%%%%%%%%%%%%%%%%%%%%%%%%%%%%%%%%%%%%%%%%%%%%%%%%%%%%%%%%%%%%%%%%%%%%%%%%%%%%%%%%
%
\subsection{The limit of vanishing dispersive self-energy}
%
%%%%%%%%%%%%%%%%%%%%%%%%%%%%%%%%%%%%%%%%%%%%%%%%%%%%%%%%%%%%%%%%%%%%%%%%%%%%%%%%

In the limit of vanishing dispersive self-energy $\Sigma^h$, the spectral function (\ref{spectral}) is 
diagonal in the mass-diagonal basis (see \cite{FHKR} for details):
\begin{equation}
 {\cal A}_{ij}(k) \equiv \pi\,{\rm sgn}(k_0) (\kdag + m_i)
            \delta(k^2 - m_i^2)\delta_{ij} \,,
\label{free_spectral}
\end{equation}
where $m_i$ are (positive and real) elements of the diagonalized mass matrix $m_d = U m V^\dagger$, 
$U$ and $V$ are unitary $N \times N$-matrices, and the mass-basis correlators and self-energies 
are defined as $S_d(k,t) \equiv Y S(k,t) X^\dagger$ and $\Sigma_d(k,t) \equiv X \Sigma(k,t) Y^\dagger$ 
with $Y \equiv P_L \otimes U + P_R \otimes V$ and $X \equiv \gamma^0 Y \gamma^0$. Using the spectral 
function (\ref{free_spectral}) in Eq.~(\ref{phase_space_eq2}) and parametrizing the most general 
(spatially homogeneous) local correlator $S^{<,>}_\mathbf{k}(t,t)$ in terms of the helicity and energy 
projection matrices:  $P_h({\bf k}) \equiv \frac12\big(1 + 
h ({\bf k} \cdot \gamma^0 \mathbf{\gamma} \gamma^5)/|{\bf k}|\big)$ where $h=\pm 1$, and $\kdag_{i\pm} + m_i$, we get
\begin{align}
iS^{<,>}_{ij}(k,t) = 
2\pi \sum_{h,\pm} \frac{1}{4\omega_i \omega_j} P_h (\kdag_{i\pm} + m_i) \Big[&(\kdag_{j\pm} + m_j) 
f^{m<,>}_{ijh\pm} \delta(k_0 \mp \bar\omega_{ij})
\nonumber\\
-\,&(\kdag_{j\mp} + m_j) f^{c<,>}_{ijh\pm} \delta(k_0 \mp \Delta\omega_{ij})\Big]\,,
\label{full_wightman_diag}
\end{align}
where $k^{\mu}_{i\pm} \equiv (\pm\omega_i,\mathbf{k})$, $\omega_{i}(\mathbf{k}) 
\equiv \sqrt{\mathbf{k}^2 + m_i^2}$, $\bar\omega_{ij} \equiv (\omega_i + \omega_j)/2$ and 
$\Delta\omega_{ij} \equiv (\omega_i - \omega_j)/2$. The Wightman functions $S^{<,>}_{ij}(k,t$) have a 
singular phase-space structure with the shells $k_0 = \pm \bar\omega_{ij}$ and 
$k_0 = \pm \Delta\omega_{ij}$ multiplied by the corresponding projection matrices and the on-shell 
distribution functions $f^{m<,>}_{ijh\pm}(\mathbf{k},t)$ and $f^{c<,>}_{ijh\pm}(\mathbf{k},t)$, which 
describe the amount of flavour coherence between the flavour eigenstates with energies $\pm \omega_i$ and
 $\pm \omega_j$, respectively. For more detailed discussion and an alternative derivation of the Wightman
 functions (\ref{full_wightman_diag}), see \cite{FHKR}.

%%%%%%%%%%%%%%%%%%%%%%%%%%
%%%%%%%%%%%%%%%%%%%%%%%%%%
\section{Kinetic equations I}
\label{sec:kinetic_eq1}
%%%%%%%%%%%%%%%%%%%%%%%%%%
%%%%%%%%%%%%%%%%%%%%%%%%%%

The first order KB-equations (\ref{KB_eq2}), written for the hermitian local correlators
$\bar S^{<,>}_\mathbf{k}(t,t) \equiv i S^{<,>}_\mathbf{k}(t,t) \gamma^0$ are given by\footnote{The hermitian conjugate terms arise from the limit $\partial_t \bar S^{<,>}_\mathbf{k}(t,t) 
= \lim_{t_2 \to t_1}\big(\partial_{t_1}\bar S^{<,>}_\mathbf{k}(t_1,t_2) + 
\partial_{t_2}\bar S^{<,>}_\mathbf{k}(t_1,t_2)\big) =  \lim_{t_2 \to t_1}
\partial_{t_1}\big(\bar S^{<,>}_\mathbf{k}(t_1,t_2) + \bar S^{<,> \dagger}_\mathbf{k}(t_1,t_2)\big)$, 
or alternatively Eq.~(\ref{kinetic_eq}) can be obtained as $k_0$-integral of the antihermitian 
part of KB-equation (\ref{KB_eq2}) in Wigner representation.}
\begin{align}
\partial_t \bar S^{<,>}_\mathbf{k}(t,t) = -i\big[H_{\mathbf{k}0}(t), \bar S^{<,>}_\mathbf{k}(t,t)\big]  
- &i \int {\rm d} t^\prime\,\big(\bar\Sigma^h_\mathbf{k}(t,t^\prime)\bar S^{<,>}_\mathbf{k}(t^\prime,t) 
- \,\bar S^{<,>}_\mathbf{k}(t, t^\prime)\bar\Sigma^h_\mathbf{k}(t^\prime,t)\big) 
\nonumber\\
\pm& \gamma^0\big({\cal C}^{\rm coll}_\mathbf{k}(t,t) + {\cal C}^{\rm coll}_\mathbf{k}(t,t)^\dagger\big)
\gamma^0\,,
\label{kinetic_eq}
\end{align}
where $\bar\Sigma^h \equiv \gamma^0 \Sigma^h$. Here we have dropped the interaction terms 
$\sim \Sigma^{<,>} S^h_{\mathbf{k}}$, which are related to finite scattering widths of the propagators
 and are therefore beyond quasiparticle approximation. The kinetic equations (\ref{kinetic_eq})
 are non-local in $t$, because the interaction terms (convolutions) involve non-local correlators 
$S^h_{\mathbf{k}}(t^\prime,t)$. Moreover, the loop expansion of the self-energies $\Sigma$ involves 
in general (non-local) correlators $S^h_{\mathbf{k}}(w_0,w_0^\prime)$, where vertex time-arguments 
$w_0$ and $w_0^\prime$ are integrated over. However, {\em local kinetic equations} can 
be obtained up to first order by using the zeroth order equations (following from Eqs.~(\ref{combined_eq}) and
(\ref{spectral_additivity})):  
\begin{equation}
S^{<,>}_\mathbf{k}(w_0,w_0^\prime) = {\cal A}_\mathbf{k}(w_0-t) {\cal A}_{\mathbf{k}0}^{-1} 
S^{<,>}_\mathbf{k}(t,t){\cal A}_{\mathbf{k}0}^{-1}{\cal A}_\mathbf{k}(t-w_0^\prime) \,,
\label{zeroth_kinetic_eq} 
\end{equation}
to expand all non-local correlators in terms of the local correlators $S^{<,>}_\mathbf{k}(t,t)$ 
to get a closure for the kinetic equations (\ref{kinetic_eq}). This expansion of the interaction 
terms can neatly be done in momentum space by enlarged set of Feynman rules involving flavour-coherent 
effective propagators.

%%%%%%%%%%%%%%%%%%%%%%%%%%
%%%%%%%%%%%%%%%%%%%%%%%%%%
\section{Momentum space Feynman rules}
\label{sec:feynman}
%%%%%%%%%%%%%%%%%%%%%%%%%%
%%%%%%%%%%%%%%%%%%%%%%%%%%

To derive the cQPA Feynman rules, we first rewrite the interaction convolutions in Eq.~(\ref{kinetic_eq}) 
to the form
\begin{align}
I(t) \equiv& \int {\rm d} t^\prime\,\Sigma_\mathbf{k}(t,t^\prime)S^{<,>}_\mathbf{k}(t^\prime,t)
\nonumber\\
=& \int {\rm d} w_0 {\rm d} w_0^\prime {\rm d} w_0^{\prime\prime}\,\delta(t-w_0)
\Sigma_\mathbf{k}(w_0,w_0^\prime)S^{<,>}_\mathbf{k}(w_0^\prime,w_0^{\prime\prime})
\delta(w_0^{\prime\prime}-t)\,.
\label{convo1}
\end{align}
Then we expand all Wightman functions in terms of $S^{<,>}_\mathbf{k}(t,t)$ using 
Eq.~(\ref{zeroth_kinetic_eq}), and perform a shift of integration variable: 
$w_0 \to \tilde w_0 = w_0 - t$, for the time-arguments $w_0, w_0^\prime, w_0^{\prime\prime}$
and all vertex time-arguments inside $\Sigma(w_0,w_0^\prime)$. As a result all 
Wightman functions in Eq.~(\ref{convo1}) are replaced by (effective) propagators 
\begin{equation}
\tilde{S}^{<,>}_{\mathbf{k}}(\tilde w_0, \tilde w_0^\prime) \equiv 
S^{<,>}_{\mathbf{k}}(\tilde w_0 + t, \tilde w_0^\prime + t) = {\cal A}_\mathbf{k}(\tilde w_0) 
{\cal A}_{\mathbf{k}0}^{-1} S^{<,>}_\mathbf{k}(t,t){\cal A}_{\mathbf{k}0}^{-1}
{\cal A}_\mathbf{k}(-\tilde w_0^\prime)\,,
\label{eff_prop_two-time}
\end{equation}
while the translation invariant pole-propagators appearing in the generic loop diagrams in Closed 
Time Path (CTP)-formalism \cite{SK-formalism} remain invariant, since 
$S^{r,a}(w_0 - w_0^\prime) = S^{r,a}(\tilde w_0 - \tilde w_0^\prime)$.
The primary CTP-propagators, given by\footnote{Identical relations hold for the self-energy components $\Sigma^{ab}$.} 
\begin{align}
S^{++}& \equiv S^t = \phantom{-}S^r - S^< \,,  &  S^{+-} \equiv -S^<\,, \phantom{hann}
\nonumber\\ 
S^{--}& \equiv S^{\bar t}   = - S^a - S^< \,, &  S^{-+} \equiv \phantom{-}S^> \,,\phantom{hann}
\label{CTP-props}
\end{align}
then transform accordingly. Similar transformation properties apply to (coherent or translation
 invariant) propagators of the other fields coupled to the field $\psi$, appearing inside the 
self-energies. By writing the effective propagators (\ref{eff_prop_two-time}) as double 
Fourier-transforms w.r.t $\tilde w_0$ and $\tilde w_0^\prime$, and performing the time integrals 
over all vertices $\tilde w_0, \tilde w_0^\prime, \tilde w_0^{\prime\prime}$ and the ones inside 
$\tilde{\Sigma}_{\mathbf{k}}(\tilde w_0, \tilde w_0^\prime) \equiv 
\Sigma_{\mathbf{k}}(\tilde w_0 + t, \tilde w_0^\prime + t)$,
we then obtain
\begin{align}
I(t) =& \int {\rm d} \tilde w_0 {\rm d} \tilde w_0^\prime {\rm d} \tilde w_0^{\prime\prime}\,
\delta(\tilde w_0)
\tilde{\Sigma}_{\mathbf{k}}(\tilde w_0,\tilde w_0^\prime)
\tilde{S}^{<,>}_{\mathbf{k}}(\tilde w_0^\prime,\tilde w_0^{\prime\prime})\delta(\tilde w_0^{\prime\prime})
\nonumber\\
=&- \int \frac{{\rm d}k_0}{2\pi }\, i \Sigma_{\rm eff}(k,t) i S^{<,>}_{\rm eff, in}(k,t)\,,
\label{int_convo}
\end{align}
where the effective momentum space self-energies are formally given by\footnote{Note that this relation is not the standard Wigner transform of the self energy.}
\begin{align}
\Sigma_{\rm eff}(k,t) \equiv& \int d \tilde w_0^\prime e^{-i k_0 \tilde w_0^\prime} \tilde{\Sigma}_{\mathbf{k}}(0,\tilde w_0^\prime)
\nonumber\\
=& \int d w_0^\prime e^{i k_0(t - w_0^\prime)} \Sigma_{\mathbf{k}}(t, w_0^\prime)\,,
\end{align}
while the effective in-propagators are defined as
\begin{equation}
i S^{<,>}_{\rm eff, in}(k,t) = {\cal A}(k){\cal A}_{\mathbf{k}0}^{-1} i S^{<,>}_\mathbf{k}(t,t)\,.
\label{eff_in-prop}
\end{equation}
The effective Wightman functions {\em inside} $\Sigma_{\rm eff}(k,t)$ are written as
\begin{equation}
i S^{<,>}_{\rm eff}(k,k^\prime,t) = {\cal A}(k) F^{<,>}_{\mathbf{k}\,\mathbf{k^\prime}}(t,t) 
{\cal A}(k^\prime)\,,
\label{eff_prop}
\end{equation}
with 
\begin{equation}
F^{<,>}_{\mathbf{k}\,\mathbf{k^\prime}}(t,t) \equiv {\cal A}_{\mathbf{k}0}^{-1} 
i S^{<,>}_\mathbf{k}(t,t) {\cal A}_{\mathbf{k}0}^{-1} \,(2\pi)^3 \delta^3(\mathbf{k}-\mathbf{k^\prime})\,.
\label{eff_vertex}
\end{equation}
Moreover, $S^{<,>}_{\rm eff}(k,k^\prime,t)$ are accompanied by an additional 4-momentum 
integral w.r.t $k^\prime$. On the other hand, the pole propagators inside $\Sigma_{\rm eff}(k,t)$ 
have the standard expressions, given by Eq.~(\ref{pole}). The effective self-energies 
$\Sigma_{\rm eff}(k,t)$ are then simply the usual (CTP) self-energies in momentum space, where 
all propagators are replaced by the corresponding effective propagators. We note that
 Eq.~(\ref{int_convo}) is of the same form as the (standard) interaction term arising from zeroth
 order truncation of the ($\Diamond$-)gradient expansion in Wigner representation (see ref.~\cite{FHKR}
 and Eqs.~(\ref{convo_wigner}-\ref{diamond}) below).
 However, Eq.~(\ref{int_convo}) with effective propagators and self-energies involves a resummation
 of rapid coherence oscillations of frequencies $\sim k$, contributing to all orders in the
 $\Diamond$-expansion. Only in the case of vanishing coherence contributions, $\Sigma_{\rm eff}(k,t)$ and
 (flavour diagonal) $S^{<,>}_{\rm eff, in}(k,t)$ reduce to standard Wigner representation expressions
 $\Sigma(k,t)$ and $S^{<,>}(k,t)$, and $I(t)$ reduces to the zeroth order truncation of the
 $\Diamond$-expansion.

\begin{figure}[t]
\centering
\includegraphics[width=0.95 \textwidth]{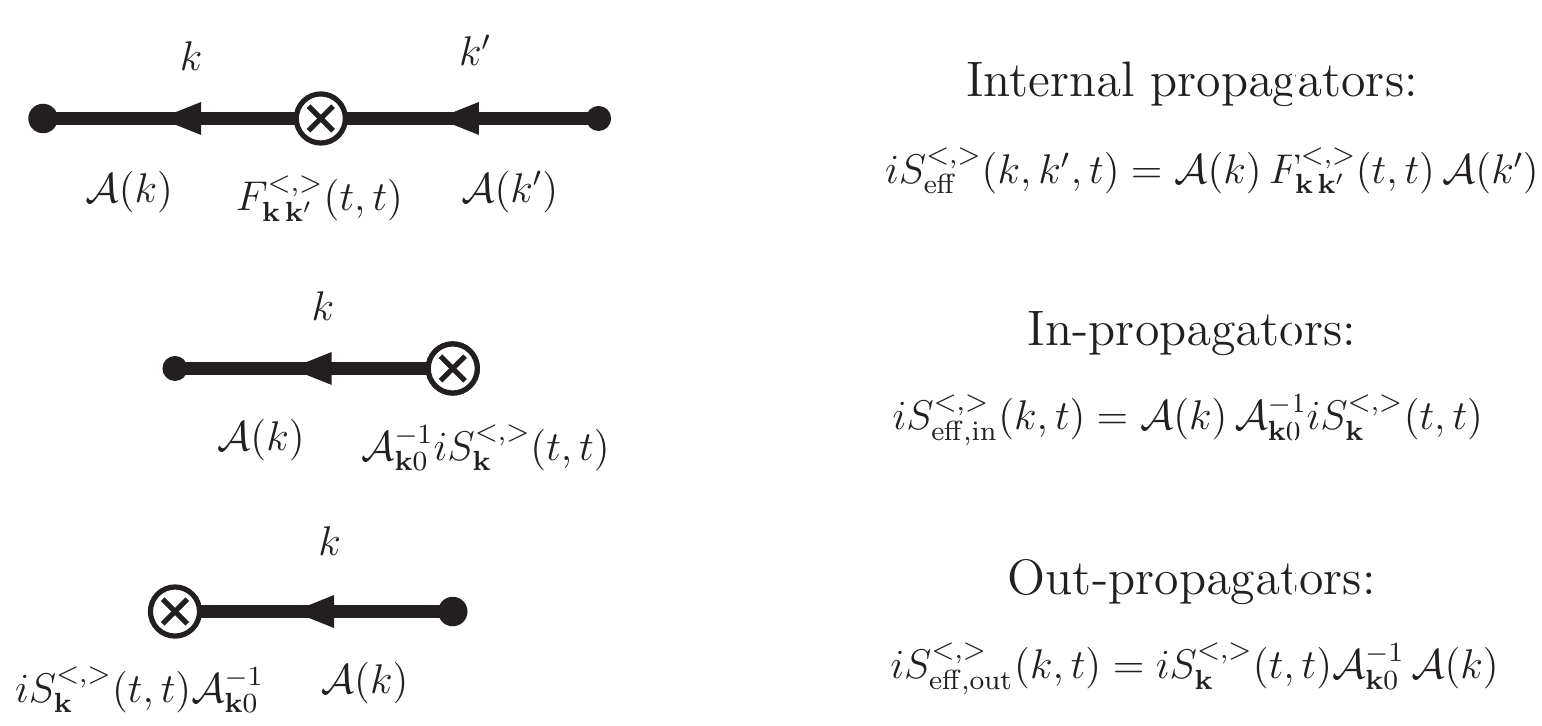}
\caption{The Feynman rules for the flavour-coherent (effective) Wightman propagators. The 
propagators are composed of the spectral function ${\cal A}(k)$ (fermionic lines) and the effective 
2-point vertices (circled crosses) encoding the quantum statistical information.}
\label{fig:FeynmanRules}
\end{figure}

Furthermore, the hermitian conjugate interaction terms in Eq.~(\ref{kinetic_eq}) are given by\footnote{Note that the effective self energies $\Sigma_{\rm eff}(k,t)$ do not follow the hermiticity properties of the Wigner transformed self-energies $\Sigma(k,t)$.}
\begin{equation}
I(t)^\dagger = \int \frac{{\rm d}k_0}{2\pi } \gamma^0 i S^{<,>}_{\rm eff, out}(k,t) \gamma^0 
i \Sigma_{\rm eff}^\dagger(k,t)\,,
\label{int_convo_conjugate}
\end{equation}
where the effective out-propagators are given by
\begin{equation}
i S^{<,>}_{\rm eff, out}(k,t) = - \gamma^0 i S^{<,> \dagger}_{\rm eff, in}(k,t) \gamma^0 
= i S^{<,>}_\mathbf{k}(t,t) {\cal A}_{\mathbf{k}0}^{-1}{\cal A}(k) \,.
\label{eff_out-prop}
\end{equation}
To summarize, we list the {\em cQPA Feynman rules} to compute the interaction terms $I(t)$ and $I^\dagger(t)$
 of Eq.~(\ref{kinetic_eq}) in momentum space in terms of (effective) flavour-coherent propagators involving
the (dynamical) local correlators $S^{<,>}_\mathbf{k}(t,t)$: (see Fig.~\ref{fig:FeynmanRules})
\begin{itemize}
\item
Draw the self-energy diagram(s) for $i \Sigma_{\rm eff}(k,t)$ and use the standard CTP-Feynman rules 
for the vertices in momentum space, and apply CTP-indices for the propagators. The momentum
conservation delta function is not applied to the out-vertex of $i \Sigma_{\rm eff}(k,t)$.\footnote{In
 comparison, by the standard (momentum space) Feynman rules, any vertex can be singled out to be
 lacking the momentum conservation delta function.}
\item
Use the relations (\ref{CTP-props}) to write the CTP-propagators in terms of $i S^{r,a}$ and $i S^{<,>}$. 
\item
For the pole propagators, use the standard expressions $i S^{r,a}(k)$ of Eq.~(\ref{pole}) (with $\Gamma \to 0$) with the usual (single) $k$-integration.
\item
For the internal Wightman propagators, use $i S^{<,>}_{\rm eff}(k,k^\prime,t)$ of Eq.~(\ref{eff_prop}) 
with double 4-momentum integration over $k$ and $k^\prime$.
\item
For the external in- and out- Wightman propagators, use $i S^{<,>}_{\rm eff, in}(k,t)$ and 
$i S^{<,>}_{\rm eff, out}(k,t)$ of Eqs.~(\ref{eff_in-prop}) and (\ref{eff_out-prop}), respectively.
\end{itemize}
\begin{figure}[t]
\centering
\includegraphics[width=0.8 \textwidth]{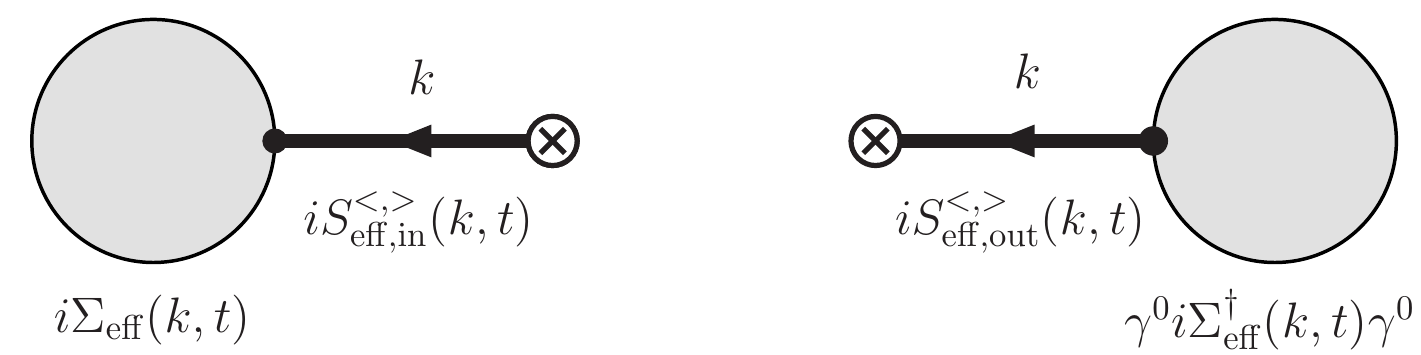}
\caption{A diagrammatic representation of the interaction terms $I(t)$ (left) and $\gamma^0 I^\dagger(t) \gamma^0$ (right) in the kinetic equation (\ref{kinetic_eq}).}
\label{fig:int_terms}
\end{figure}
In figure \ref{fig:FeynmanRules} we show diagrammatic representations of the flavour-coherent cQPA 
propagators, which are composed of the spectral function ${\cal A}(k)$ (\ref{spectral}) and the 
{\em effective 2-point vertices} $F^{<,>}_{\mathbf{k}\,\mathbf{k^\prime}}(t,t)$ (\ref{eff_vertex}) 
(or ${\cal A}_{\mathbf{k}0}^{-1} i S^{<,>}_\mathbf{k}(t,t)$ and 
$i S^{<,>}_\mathbf{k}(t,t){\cal A}_{\mathbf{k}0}^{-1}$ for in/out-propagators, respectively), encoding 
the quantum statistical information. In figure \ref{fig:int_terms} we show generic structures of the 
interaction terms $I(t)$ and $I^\dagger(t)$ appearing in the kinetic equation (\ref{kinetic_eq}).

These Feynman rules can be used to compute any type of Green's function in CTP formalism, not only the 
self-energies appearing in (KB-) kinetic equations. For example, a perturbative contribution to full 
Wightman function becomes:
\begin{align}
iS^{<}_{\rm pert}(k,t) =& \int {\rm d} (t_1-t_2) e^{i k_0 (t_1-t_2)}\int {\rm d} w_0 {\rm d} w_0^\prime\,
iS^{<}_{\mathbf{k}}(t_1,w_0)
i\Sigma^{>}_{\mathbf{k}}(w_0,w_0^\prime)
iS^{<}_{\mathbf{k}}(w_0^\prime,t_2) + \ldots
\nonumber\\
=& 2 \int \frac{{\rm d}k_0^\prime}{2\pi } \frac{{\rm d}^4 p}{(2\pi)^4 } 
iS^{<}_{{\rm eff}}\big((k_0^\prime,\mathbf{k}),p,t\big)i\Sigma^{>}_{\rm eff}(p,t) 
iS^{<}_{\rm eff}\big(p,(2 k^0 - k_0^{\prime},\mathbf{k}),t\big) + \ldots\,,
\label{pert_corr1}
\end{align}
where we have again used shifts in the integration variables such as 
$w_0 \to \tilde w_0 = w_0 - t$.
We notice the similar convolution structure as in the case of zeroth order correlator in 
Eq.~(\ref{phase_space_eq2}). For the corresponding local correlator we then get by $k_0$-integration 
over Eq.~(\ref{pert_corr1}):
\begin{equation}
i S^{<}_{\mathbf{k},{\rm pert}}(t,t) = \int \frac{{\rm d}k_0}{2\pi } 
i S^{<}_{{\rm eff, out}}(k,t) i \Sigma^{>}_{\rm eff}(k,t) 
i S^{<}_{\rm eff, in}(k,t) + \ldots\,.
\label{pert_corr_local}
\end{equation}
The perturbative expansion in Eqs.~(\ref{pert_corr1}-\ref{pert_corr_local}) is performed around 
the time instant $t$, \ie the implicit local correlators $S^{<,>}_\mathbf{k}(t,t)$ appearing on the RHS
 carry the full (quantum) statistical information. Therefore Eq.~(\ref{pert_corr_local}) is actually an
 equation for $S^{<}_\mathbf{k}(t,t)$. On the other hand, by performing the perturbative expansion
 around the ``initial'' local correlators $S^{<,>}_\mathbf{k}(0,0)$, we find a relation
\begin{equation}
i S^{<}_{\rm pert}(k,t) = {\cal A}_{\mathbf{k}}(t){\cal A}_{\mathbf{k}0}^{-1} i S^{<}_{\rm pert}(k,0)
{\cal A}_{\mathbf{k}0}^{-1}{\cal A}_{\mathbf{k}}(-t)\,,
\label{pert_corr2}
\end{equation}
where $i S^{<}_{\rm pert}(k,0)$ can be read off Eq~(\ref{pert_corr1}) by setting $t = 0$. 
By identifying
 $U_\mathbf{k}(t) = {\cal A}_\mathbf{k}(y){\cal A}_{\mathbf{k}0}^{-1}$ as a quasiparticle time-evolution
 operator according to Eq.~(\ref{time-evo_operator}), we see that Eq.~(\ref{pert_corr2})
 corresponds to perturbative time evolution in {\em quasiparticle interaction picture}. It is well known that
 out-of-equilibrium perturbative time evolution suffers from {\em secularity}, and the error becomes 
large for $\Gamma t \gtrsim 1$ (see {\em e.g.} \cite{Berges04}). In contradistinction, in cQPA we do not use the 
perturbative expansion around initial time correlators. Instead we expand the interaction terms $I(t)$ in the kinetic equation 
(\ref{kinetic_eq}) in terms of the local correlators $S^{<,>}_\mathbf{k}(t,t)$, 
and solve the resulting local time-evolution equations for $S^{<,>}_\mathbf{k}(t,t)$. 
Since the expansion is made around the time-evolved correlators, this
procedure does {\em not} suffer from secular error growth.
We will show this in section \ref{sec:error} below by computing iteratively the magnitude 
of the neglected correction terms.

%%%%%%%%%%%%%%%%%%%%%%%%%%%%%%%%%%%%%%%%%%%%%%%%%%%%%%%%%%%%%%%%%%%%%%%%%%%%%%%%
%
\subsection{Example: One-loop self-energy in Yukawa theory}
%
%%%%%%%%%%%%%%%%%%%%%%%%%%%%%%%%%%%%%%%%%%%%%%%%%%%%%%%%%%%%%%%%%%%%%%%%%%%%%%%%

\begin{figure}[t]
\centering
\includegraphics[width=0.95 \textwidth]{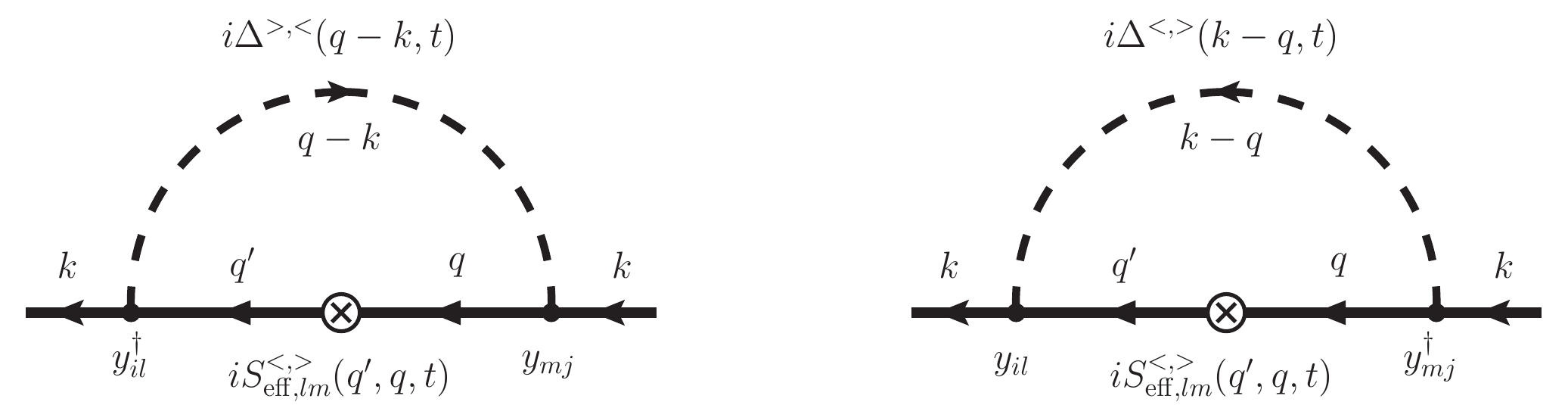}
\caption{The Feynman diagrams contributing to the (effective) self-energies $\Sigma_{{\rm eff},ij}^{<,>}(k,t)$ at one-loop level ($i,j,l,m$ are flavour indices).}
\label{fig:one-loop}
\end{figure}

As an example of the Feynman rules, we write down the effective one-loop self-energies
 $i \Sigma_{\rm eff}^{<,>}(k,t)$ in a Yukawa theory described by the (interaction) Lagrangian 
\begin{equation} 
{\cal L}_{\rm int} = - y_{ij}\; \bar \psi_i\, \phi \, P_L \psi_j + h.c. \,,
\label{interaction}
\end{equation}
where $y_{ij}$ is a complex Yukawa coupling matrix in flavour indices. The one-loop self-energy diagram(s)
 can be obtained for example from the two-particle irreducible (2PI) effective action \cite{2PI} two-loop
 vacuum diagram:
\begin{equation} 
\Gamma_{\rm 2PI} = - y_{ij} y_{lm}^\dagger \int_C {\rm d}^4u\, {\rm d}^4v\, {\rm Tr}\left[P_L S_{jl}(u,v) P_R S_{mi}(v,u)\right]\Delta(u,v) \,,
\label{gamma2pI}
\end{equation}
where sum over repeated flavour indices $i,j,l,m$ is understood, and the integration is over the Keldysh 
path (see {{\em e.g.}}~\cite{HKR1}). $\Delta(u,v)$ denotes the scalar field (CTP) propagator. The fermion 
self-energy now follows by a functional differentiation:
\begin{align}
   \Sigma^{ab}_{ij}(u,v) =&  -iab \frac{\delta \Gamma_2[S]}{\delta S^{ba}_{ji}(v,u)} 
\nonumber\\
=& i \left[ y_{il}^\dagger y_{mj} P_R S^{ab}_{lm}(u,v) P_L \Delta^{ba}(v,u) + y_{il} y_{mj}^\dagger P_L S^{ab}_{lm}(u,v) P_R \Delta^{ab}(u,v)\right]\,.
\label{sigma-ab2}
\end{align}   
By performing a Wigner transformation, we now find for the self-energies $i \Sigma^{<,>}(k,t)$
\begin{align} 
i\Sigma^{<,>}_{ij}(k,t) = \int \frac{{\rm d}^4q}{(2\pi)^4} \frac{{\rm d}^4p}{(2\pi)^4} 
(2\pi )^4 \delta^4(k - q - p) 
\big[&y_{il}^\dagger y_{mj} P_R iS^{<,>}_{lm}(q,t) P_L i\Delta^{>,<}(-p,t)
\nonumber\\
+& y_{il} y_{mj}^\dagger P_L iS^{<,>}_{lm}(q,t) P_R i\Delta^{<,>}(p,t) \big]\,.
\label{wignerself}
\end{align}
On the other hand, the effective self-energies $i \Sigma^{<,>}_{\rm eff}(k,t)$ 
appearing in the kinetic equations (\ref{kinetic_eq}) (and below in 
Eqs.~(\ref{kinetic_eq2}) and (\ref{kinetic_eq3})) are obtained by using the 
cQPA Feynman rules developed in this section. By assuming that the scalar 
propagators $\Delta^{<,>}$ are non-coherent,\footnote{The single flavour 
scalar field could also mediate (flavour diagonal) 
particle-antiparticle coherence (see refs.~\cite{HKR3,HKR4}). The non-coherent 
scalar propagators $\Delta^{<,>}(p,t)$ correspond to the usual quasiparticle 
propagators (see {\em e.g.} \cite{QPA_props,sum-rule_QPA}), possibly involving 
out-of-equilibrium particle number distribution functions.} we get 
(see Fig.~\ref{fig:one-loop})
\begin{align} 
i\Sigma^{<,>}_{{\rm eff},ij}(k,t) 
=  \int \frac{{\rm d}^4q}{(2\pi)^4} 
      \frac{{\rm d}^4q^\prime}{(2\pi)^4} 
      \frac{{\rm d}^4p}{(2\pi)^4} 
      &(2\pi )^4 \delta^4(k - q - p) 
\nonumber\\
\times\big[&y_{il}^\dagger y_{mj} P_R iS^{<,>}_{{\rm eff},lm}(q^\prime,q,t) P_L i\Delta^{>,<}(-p,t)
\nonumber\\[1mm]
+ &y_{il} y_{mj}^\dagger P_L iS^{<,>}_{{\rm eff},lm}(q^\prime,q,t) P_R i\Delta^{<,>}(p,t) \big]\,.
\label{effself}
\end{align}
We notice that the 4-momentum is not formally conserved in the out-vertex of this self-energy diagram
 (unless $q^\prime = q$). However, the $q^\prime$-integration can always be performed trivially because
 of the singularity structure of $S^{<,>}_{{\rm eff}}(q^\prime,q,t)$. This results to appropriate
 energy projections of the spinor structure, such that the kinematical interpretation of the (coherent)
 collision process remains natural. By using the explicit expressions (\ref{eff_prop}-\ref{eff_vertex})
 one can write Eq.~(\ref{effself}) in terms of the local correlators $S^{<,>}_{\mathbf{q}}(t,t)$, and
 perform some of the momentum integrations. Further simplifications of Eq.~(\ref{effself}) would require
 a specification of the fermionic spectral function (\ref{spectral}) and the scalar correlators
 $\Delta^{<,>}(p,t)$, and are therefore model dependent. 
%

%%%%%%%%%%%%%%%%%%%%%%%%%%
%%%%%%%%%%%%%%%%%%%%%%%%%%
\section{Kinetic equations II}
\label{sec:kinetic_eq2}
%%%%%%%%%%%%%%%%%%%%%%%%%%
%%%%%%%%%%%%%%%%%%%%%%%%%%

In this section, we use the cQPA Feynman rules to rewrite and simplify the interaction terms in the first
 order kinetic equations (\ref{kinetic_eq}) in order to obtain (local) time-evolution equations for
 the local correlators $\bar S^{<}_\mathbf{k}(t,t)$. In the case of vanishing dispersive self-energy
 $\Sigma^h$, we also write down the corresponding flavoured quantum Boltzmann equations for the on-shell
 distribution functions, given by Eq.~(\ref{full_wightman_diag}). Finally, we present an estimation for
 the error in time evolution resulting from our approximations.
  
To begin, we use the results (\ref{int_convo}) and (\ref{int_convo_conjugate}) with 
Eqs.~(\ref{eff_in-prop}) and (\ref{eff_out-prop}) for the interaction terms in the kinetic equation 
(\ref{kinetic_eq}) to get for $S^<$
\begin{align}
\partial_t \bar S^{<}_\mathbf{k}(t,t) = -i\big[\tilde H_{\mathbf{k},{\rm eff}}(t), 
\bar S^{<}_\mathbf{k}(t,t)\big]^m - \frac{1}{2}\left(\big\{\tilde \Sigma^{>}_{\mathbf{k}}(t), 
\bar S^{<}_\mathbf{k}(t,t)\big\}^m - \big\{\tilde \Sigma^{<}_{\mathbf{k}}(t), 
\bar S^{>}_\mathbf{k}(t,t)\big\}^m \right)\,,
\label{kinetic_eq2}
\end{align}
where we have defined
\begin{equation}
\tilde H_{\mathbf{k},{\rm eff}}(t) \equiv H_{\mathbf{k}0}(t) + \tilde\Sigma^h_\mathbf{k}(t)\,,
\end{equation}
and (for all components of self-energy)
\begin{align}
\tilde\Sigma_\mathbf{k}(t) \equiv \int \frac{{\rm d}k_0}{2\pi }\,\bar\Sigma_{\rm eff}(k,t)
{\cal A}(k){\cal A}_{\mathbf{k}0}^{-1}
%\nonumber\\ 
= \sum_i \bar\Sigma_{\rm eff}(\tilde\omega_{i\mathbf{k}},
\mathbf{k},t) A_{i \mathbf{k}}{\cal A}_{\mathbf{k}0}^{-1}\,,
\label{tilde_Sigma}
\end{align}
where we denote $\bar \Sigma^h \equiv \gamma^0 \Sigma^h$ and
 $\bar \Sigma^{<,>} \equiv \gamma^0 i\Sigma^{<,>}$ and we have used Eq.~(\ref{spectral}) for the
 spectral function. The generalized (anti)commutators (with respect to both Dirac and flavour indices)
 are defined as
\begin{equation}
\big[A , B \big]^m  \equiv A B  - B^\dagger A^\dagger \,, 
\quad \{A , B \}^m  \equiv A B  + B^\dagger A^\dagger \,.
\label{gen_comm}
\end{equation}
We don't need to consider the kinetic equation for $S^{>}_\mathbf{k}(t,t)$, because the Wightman functions are in general related by
\begin{equation}
S^> = -S^< - 2i{\cal A}
\label{wightman_rel}
\end{equation}
and the spectral function ${\cal A}$, given by Eq.~(\ref{spectral}), is non-dynamical.
The effective self-energies $\Sigma_{\rm eff}(k,t)$ are written in terms of the local correlators 
$\bar S^{<}_\mathbf{k}(t,t)$ (and similarly for other fields coupling to field $\psi$) using the Feynman
rules of the previous section. The kinetic equation (\ref{kinetic_eq2}) is therefore
local in time $t$ for arbitrary interactions, and hence an ordinary (coupled) first order differential 
equation in contrast to integro-differential equations with full memory integrals (see {\em e.g.}~\cite{full_memory}).
 However, the kinetic equation (\ref{kinetic_eq2}) is not in general diagonal in momentum $|\mathbf{k}|$
 due to (possible) self-interactions. Furthermore, if $\Sigma^h$ does not depend on the
 dynamical correlators, we find to leading order in $\Sigma^h$:
 $\tilde\Sigma^h_\mathbf{k}(t) = \int {\rm d} y \, \bar\Sigma^h_{\mathbf{k}}(y,t) e^{i H_{\mathbf{k}0}(t) y}$ 
such that $\tilde H_{\mathbf{k},{\rm eff}}(t) = H_{\mathbf{k},{\rm eff}}(t)$, where the adiabatic 
$t$-dependence is now explicitly shown. We have assumed this non-dynamical $\Sigma^h$ generally in
 the zeroth order equations in order to solve the phase-space structure of the cQPA propagators.
 However, in the (first order) kinetic equation (\ref{kinetic_eq2}) we can relax this assumption and
 consider dispersive self-energies $\Sigma^h$ which involve dynamical correlators and (flavour) coherence
 in the same footing as absorptive self-energies $\Sigma^{<,>}$.        

In the case of an interaction with a thermal bath, the leading order self-energies do not depend on
 the dynamical correlators (thus $\Sigma_{\rm eff}(k,t) \to \Sigma(k,t)$) and they satisfy KMS relations \cite{KMS} 
($\beta \equiv 1/T$):
\begin{equation}
\Sigma^>(k,t) = e^{\beta k_0} \Sigma^<(k,t)\,.
\label{KMS}
\end{equation}
Using the relations (\ref{wightman_rel}-\ref{KMS}) in the kinetic equation (\ref{kinetic_eq2}),
 we then find a linear equation
\begin{align}
\partial_t \bar S^{<}_\mathbf{k}(t,t) = -i\big[ H_{\mathbf{k},{\rm eff}}(t), 
\bar S^{<}_\mathbf{k}(t,t)\big]^m - \Big(\big\{\tilde\Gamma_{\mathbf{k}}(t), 
\bar S^{<}_\mathbf{k}(t,t)\big\}^m - \int \frac{{\rm d}k_0}{2\pi }\,\big\{\bar\Gamma(k,t), 
\bar S^{<}_{\rm eq}(k,t)\big\}^m \Big)\,,
\label{kinetic_eq3}
\end{align}
where $\tilde\Gamma_{\mathbf{k}}(t)$ is given by the analogous relation to Eq.~(\ref{tilde_Sigma}) 
with $\bar\Gamma = \gamma^0 \Gamma = (\bar\Sigma^> + \bar\Sigma^<)/2$, and 
\begin{align}
iS^{<}_{\rm eq}(k,t) =& 2 f_{\rm eq}(k_0) A(k,t)
\nonumber\\ 
=& 2\pi\,{\rm sgn}(k_0)f_{\rm eq}(k_0)
\delta\big(\kdag - m_h(t) - i \gamma^5 m_a(t) - \Sigma^h(k,t)\big)\,,
\end{align}
with $f_{\rm eq}(k_0) = (e^{\beta k_0} + 1)^{-1}$, is the standard quasiparticle Wightman function in 
(local) thermal equilibrium \cite{QPA_props}.

The structure of the kinetic equation (\ref{kinetic_eq2}) or (\ref{kinetic_eq3}) with generalized
(anti)commutators is very similar to the equations of motion for the
flavoured density matrix in the standard density matrix approach to flavour mixing phenomena
(see {\em e.g.}~ref.~\cite{neutrino_osc1}).  Indeed, the local correlator $\bar S^{<}_{\mathbf{k},ij}(t,t)$  
is closely related to the density matrix in the basis of eigenstates
$u^*_i(\mathbf{k},h) a^\dagger_{i \mathbf{k}}| 0 \rangle$ and 
$v_i(\mathbf{k},h) b^\dagger_{i \mathbf{k}}| 0 \rangle$.
However, to obtain the usual equation of motion for the integrated density matrix 
in the flavour indices only, one would need to make a  
factorization ansatz in Eq.~(\ref{kinetic_eq2}):
\begin{equation}
\bar S^{<}_{\mathbf{k},ij}(t,t) \sim g(\mathbf{k}) \rho_{ij}(t)\,, 
\end{equation}
where the dependence on the momentum $\mathbf{k}$ and Dirac-indices is factorized to the function 
$g(\mathbf{k})$. In general, this kind of ansatz is not justified unless some stringent extra constraints
are imposed.  We want to emphasize that our kinetic
equations (\ref{kinetic_eq2}) and (\ref{kinetic_eq3}) with the self-energy functions 
(\ref{tilde_Sigma}) are derived from first principles using the methods of nonequilibrium 
quantum field theory. 

In order to solve the kinetic equation (\ref{kinetic_eq2}) for $S^{<}_\mathbf{k}(t,t)$, one needs to
 parametrize the most general spatially homogeneous local correlator $S^{<}_\mathbf{k}(t,t)$
 in terms of eight independent matrices of (sub-)Dirac algebra, consisting of the products of $\gamma^0$,
 $\mathbf{k}\cdot\mathbf{\gamma}$ and $\gamma^5$. Then, using the hermiticity properties (under the Dirac
 and flavour indices) we obtain coupled equations of motion for $8 N^2$ real functions of $|\mathbf{k}|$
 and $t$, where $N$ is the number of flavours. We note that the equation (\ref{kinetic_eq2}) can be solved
 in {\em any} flavour basis, where the quasiparticle spectral function (\ref{spectral}) is determined.

%%%%%%%%%%%%%%%%%%%%%%%%%%%%%%%%%%%%%%%%%%%%%%%%%%%%%%%%%%%%%%%%%%%%%%%%%%%%%%%%
%
\subsection{Flavoured quantum Boltzmann equations}
%
%%%%%%%%%%%%%%%%%%%%%%%%%%%%%%%%%%%%%%%%%%%%%%%%%%%%%%%%%%%%%%%%%%%%%%%%%%%%%%%%

In the limit of vanishing dispersive self-energy $\Sigma^h$, we can use Eq.~(\ref{full_wightman_diag})
 for the Wightman functions in the mass basis, where the on-shell distribution functions
 $f^{m<,>}_{ijh\pm}(\mathbf{k},t)$ and $f^{c<,>}_{ijh\pm}(\mathbf{k},t)$ are the dynamical variables.
 Using the projection operators $P_h$ and $\kdag_{i\pm} + m_i$, we can then solve the distribution
 functions in terms of local correlators $S^{<,>}_\mathbf{k}(t,t)$ and use the kinetic equation
 (\ref{kinetic_eq2}) to obtain
\begin{align}
 \partial_t f^{m<,>}_{ijh\pm}  
  =&  \mp i 2 \Delta \omega_{ij} f^{m<,>}_{ijh\pm }   
 + i X^m_{h\pm}[f]_{ij} + {\cal C}^m_{h\pm}[f]_{ij} \,,
\label{eom_on-shell1}
\\[2mm]
\partial_t f^{c<,>}_{ijh\pm} 
  =&  \mp i 2 \bar \omega_{ij} f^{c<,>}_{ijh\pm }  
 + i X^c_{h\pm}[f]_{ij} + {\cal C}^c_{h\pm}[f]_{ij} \,, 
\label{eom_on-shell2}
\end{align}
where $X^{m,c}_{h\pm}[f]$ involve terms proportional to mass and mixing gradients: 
$\partial_t m_i / \omega_i^2$ and 
$\Xi'^{\pm} \equiv i\big(V\partial_tV^\dagger \pm U\partial_tU^\dagger\big)/2$, and 
${\cal C}^{m,c}_{h\pm}[f]$ are the collision integrals. From these {\em flavoured quantum Boltzmann equations}
 (\ref{eom_on-shell1}-\ref{eom_on-shell2}) we can directly see that to zeroth order the on-shell functions
 $f^{m<,>}_{ijh\pm}$ and $f^{c<,>}_{ijh\pm}$ are oscillating at frequencies $\Delta \omega_{ij}$ and
 $\bar \omega_{ij}$, respectively. We present the full expressions of $X^x_{h\pm}$ and ${\cal C}^x_{h\pm}$
 in \cite{FHKR} for a general $N \times N$-mixing scenario and a general Dirac decomposition
 of spatially homogeneous self-energies.

%%%%%%%%%%%%%%%%%%%%%%%%%%%%%%%%%%%%%%%%%%%%%%%%%%%%%%%%%%%%%%%%%%%%%%%%%%%%%%%%
%
\subsection{Estimation of error in time evolution}
\label{sec:error}
%
%%%%%%%%%%%%%%%%%%%%%%%%%%%%%%%%%%%%%%%%%%%%%%%%%%%%%%%%%%%%%%%%%%%%%%%%%%%%%%%%

In order to obtain the local evolution equation (\ref{kinetic_eq2}) for $S^{<,>}_\mathbf{k}(t,t)$ 
we have extensively used the zeroth order equations (\ref{spectral_additivity}) and (\ref{combined_eq}) 
{\em inside} the interaction terms $I(t)$ of the full kinetic equation (\ref{kinetic_eq}). 
In this section we estimate the magnitude of the error resulting from this approximation.

First, we notice that the convolutions $I(t)$ can be written in Wigner representation as
\begin{align}
I_{\rm full}(t) \equiv& \int {\rm d} t^\prime\,\Sigma_\mathbf{k}(t,t^\prime)S^{<,>}_\mathbf{k}(t^\prime,t)
\nonumber\\
=& \int \frac{{\rm d}k_0}{2\pi } e^{-i\Diamond}\{ \Sigma(k,t) \}\{ S^{<,>}(k,t) \}
\nonumber\\
=& \int \frac{{\rm d}k_0}{2\pi } \sum_{n=0}^\infty \frac{1}{n!}\Big(\frac{i}{2}\Big)^n 
\partial_t^n \Big( \big( \partial_{k_0}^n \Sigma(k,t))\,S^{<,>}(k,t) \Big)\,, 
\label{convo_wigner}
\end{align}
where the $\Diamond$-operator is defined as
\begin{equation}
\Diamond\{f\}\{g\} = \frac{1}{2}\big(
                   \partial_tf \,\partial_{k_0} g
                 - \partial_{k_0} f \,\partial_t g \big)\,.
\label{diamond}
\end{equation}
Next, by using Eqs.~(\ref{phase_space_eq2}) and (\ref{kinetic_eq}) or alternatively KB-equation (\ref{KB_eq2}) 
in Wigner representation, we can parametrically estimate the time-derivatives appearing in 
Eq.~(\ref{convo_wigner}) as
\begin{align}
\partial_t S(k,t) \sim k_0 S(k,t) + \epsilon\,k_0 S(k,t)\,,
\nonumber\\
\partial_t \Sigma (k,t) \sim k_0 \Sigma(k,t) + \epsilon\,k_0 \Sigma(k,t)\,,
\label{evo_approx}
\end{align}
with 
\begin{equation}
\epsilon \sim \Big\{ \frac{\partial_t m}{k_0^2}, \frac{\Sigma(k,t)}{k_0} \Big\}\,.
\end{equation}
The zeroth ($\epsilon$-)order terms in the RHS of Eqs.~(\ref{evo_approx}) arise from the commutators 
of the (coherent) correlators $S^{<,>}$ with the Hamiltonian $H_{\mathbf{k}0}$ in Eq.~(\ref{kinetic_eq}), 
and accordingly for the self-energies involving coherent correlators. We may further estimate 
$\partial_{k_0}\Sigma(k,t) \sim \Sigma(k,t) / k_0$ to get
\begin{equation}
\partial_t \Big( \big( \partial_{k_0}\Sigma(k,t))\,S^{<,>}(k,t) \Big) 
\sim \Sigma(k,t) S(k,t) + \epsilon\,\Sigma(k,t) S(k,t)\,,
\end{equation}
and furthermore by using Eq.~(\ref{convo_wigner})
\begin{align}
I_{\rm full}(t) \sim \int \frac{{\rm d}k_0}{2\pi } \Big[ \Sigma_{\rm eff}(k,t) S^{<,>}_{\rm eff, in}(k,t) 
+ \sum_{n=1}^\infty \epsilon^n \,\Sigma(k,t) S(k,t) \Big]\,, 
\label{int_convo_full}
\end{align}
where we have noticed that all zeroth order terms have been accounted for in the effective propagators
 and self-energies. We see that the corrections to the leading order terms included in kinetic equations
 (\ref{kinetic_eq2}) are suppressed by powers of $\partial_t m / k_0^2$ and $\Sigma / k_0$. Therefore,
 we conclude that in the case of weak interactions and small gradients of the background field
 ($\epsilon \ll 1$) the error of expanding the interaction terms $I(t)$ in terms of local correlators
 $S^{<,>}_\mathbf{k}(t,t)$ by using the cQPA Feynman rules is small for any instant of time $t$.
 Consequently, the error of time evolution is not growing as $\Gamma t$ as in the case of secular evolution,
 but rather, our error is that the scattering rate $\Gamma$ (and the thermalization time $\sim 1/\Gamma$) is misevaluated
 by a small correction: $\Gamma_{\rm full} \approx (1 + \epsilon) \Gamma$.

%%%%%%%%%%%%%%%%%%%%%%%%%%%%%%%%%%%%%%%%%%%%%%%%%%%%%%%%%%%%%%%%%%%%%%%%%%%%%%%%
%%%%%%%%%%%%%%%%%%%%%%%%%%%%%%%%%%%%%%%%%%%%%%%%%%%%%%%%%%%%%%%%%%%%%%%%%%%%%%%%
%
\section{Conclusions and outlook}
\label{sec:conclusions}
%
%%%%%%%%%%%%%%%%%%%%%%%%%%%%%%%%%%%%%%%%%%%%%%%%%%%%%%%%%%%%%%%%%%%%%%%%%%%%%%%%
%%%%%%%%%%%%%%%%%%%%%%%%%%%%%%%%%%%%%%%%%%%%%%%%%%%%%%%%%%%%%%%%%%%%%%%%%%%%%%%%
%
In this work we have presented a flavour-covariant formulation of the coherent propagators and 
Feynman rules for flavour-mixing fermionic fields in spatially homogeneous and isotropic 
systems. Our formalism is based on the coherent quasiparticle approximation (cQPA) 
\cite{HKR1,HKR2,HKR3,Glasgow,Thesis_Matti,HKR4,FHKR}, where nonlocal coherence information is 
encoded in new spectral solutions at off-shell momenta. In addition to more 
simplified covariant formulation, we have generalized our previous work \cite{FHKR} 
to the case of nonzero dispersive self-energy, which leads to a broader range of applications.

The new formulation is based on the two-time representation of the correlators and it immediately 
reveals the composite nature of the cQPA Wightman function as a product of two spectral functions 
and a local correlator, interpreted as an effective two-point interaction vertex, which
 contains all quantum statistical and coherence information 
(see Eqs.~(\ref{phase_space_eq}-\ref{phase_space_eq2}) and (\ref{eff_prop})). 
This derived form can be viewed as a generalization of the well known Kadanoff-Baym ansatz to 
coherent systems. In section \ref{sec:feynman}, we have derived the cQPA Feynman rules involving the 
flavour-coherent (composite) propagators, presented in Fig.~\ref{fig:FeynmanRules}. These rules are 
designed for computing the interaction terms of the Kadanoff-Baym equations (\ref{kinetic_eq}) in 
momentum space in terms of the local correlators $S^{<,>}_\mathbf{k}(t,t)$, and they can in principle 
be applied to arbitrarily complex self-energies involving flavour-coherent propagators inside the 
loop-integrals.  

We have used the cQPA Feynman rules to derive flavoured kinetic equations (\ref{kinetic_eq2}) for 
the local correlators $S^{<,>}_\mathbf{k}(t,t)$ in the presence of dispersive and absorptive 
interactions. These equations are local in time-variable $t$, and they can as well
be rewritten as flavoured quantum Boltzmann equations for the (enlarged) set of on-shell distribution
functions $f(\mathbf{k},t)$. The structure of Eqs.~(\ref{kinetic_eq2}) is very intuitive, consisting of the 
generalized commutators and anticommutators in both flavour and Dirac indices. They are 
reminiscent of the equations of motion for the (flavoured) density matrix used {\em e.g.} in 
refs.~\cite{neutrino_osc1}. We want to 
emphasize that in our approach all the interaction terms, given by Eq.~(\ref{tilde_Sigma}), are derived 
from first principles of nonequilibrium quantum field theory by consistent approximations using the 
cQPA scheme. We have further shown that the corrections to our results are (parametrically) suppressed 
by the powers of $\partial_t m / k_0^2$ and $\Sigma(k,t)/k_0$. 

In the present work we have extended the cQPA scheme to include dispersive self-energy corrections, 
which allows the treatment of quasi-excitations in plasmas. The natural applications of 
the formalism include for example neutrino flavour oscillations in the early universe \cite{HKR6} and  
electroweak baryogenesis, where the formalism needs to be generalized to the case of a stationary planar 
symmetry, as discussed in ref.~\cite{HKR1}. Another interesting application to our formalism is a 
consistent approach to resonant leptogenesis, where the flavour coherence effects of the singlet 
neutrinos are expected to play an important role in the dynamics. In the regime of nearly degenerate 
neutrino masses, the formalism should be extended to include appropriate finite width corrections 
for the propagators.

%%%%%%%%%%%%%%%%%%%%%%%%%%%%%%%%%%%%%%%%%%%%%%%%%%%%%%%%%%%%%%%%%%%%%%%%%%%%%%%%
%%%%%%%%%%%%%%%%%%%%%%%%%%%%%%%%%%%%%%%%%%%%%%%%%%%%%%%%%%%%%%%%%%%%%%%%%%%%%%%%
%
\section*{Acknowledgments}
%
%%%%%%%%%%%%%%%%%%%%%%%%%%%%%%%%%%%%%%%%%%%%%%%%%%%%%%%%%%%%%%%%%%%%%%%%%%%%%%%%
%%%%%%%%%%%%%%%%%%%%%%%%%%%%%%%%%%%%%%%%%%%%%%%%%%%%%%%%%%%%%%%%%%%%%%%%%%%%%%%%

This work is supported by the Alexander von Humboldt Foundation and by the Gottfried Wilhelm 
Leibniz programme of the Deutsche Forschungsgemeinschaft.

%
%---------------------------- Journals----------------------------------
%

\end{document}